\documentclass[twocolumn,superscriptaddress,showpacs,amsmath,amssymb,aps,prl]{revtex4-1}

\usepackage[utf8x]{inputenc}
\usepackage{graphicx}% Include figure files
\usepackage{bm}% bold math
\usepackage{verbatim}
\usepackage{amssymb, amsmath}
\newcommand{\dd}[1]{\mathrm{d}#1\,}
\newcommand{\avg}[1]{\left\langle{#1}\right\rangle}
\renewcommand{\Re}{\mathop{\mathrm{Re}}}
\renewcommand{\Im}{\mathop{\mathrm{Im}}}

\newcommand{\tr}{\mathop{\mathrm{tr}}}

\newcommand{\DD}[1]{\mathrm{\mathbf{D}}[#1]}

\begin{document}
\title{Absorption of heat into a superconductor-normal metal-superconductor junction from a fluctuating environment}
\author{J.~Voutilainen}
\affiliation{Low Temperature Laboratory, Aalto University, P.O. Box 15100, FI-00076 AALTO, Finland}

\author{P.~Virtanen}
\affiliation{Institute for Theoretical Physics and Astrophysics, University of W\"urzburg, D-97074 W\"urzburg, Germany}

\author{T.T.~Heikkil\"a}
\affiliation{Low Temperature Laboratory, Aalto University, P.O. Box 15100, FI-00076 AALTO, Finland}

\date{\today}

\begin{abstract}
We study a diffusive superconductor-normal metal-superconductor junction in an environment with intrinsic incoherent fluctuations which couple to the junction through an electromagnetic field. When the temperature of the junction differs from that of the environment, this coupling leads to an energy transfer between the two systems, taking the junction out of equilibrium. We describe this effect in the linear response regime and show that the change in the supercurrent induced by this coupling leads to qualitative changes in the current-phase relation and for a certain range of parameters, an increase in the critical current of the junction. Besides normal metals, similar effects can be expected also in other conducting weak links.
\end{abstract}

\pacs{74.45.+c, 74.25.N-, 74.50.+r}
\maketitle

Superconducting Josephson junctions are non-linear circuit elements and therefore their state and the supercurrent carried through them depends sensitively on the properties of the field driving them. Besides the average current or voltage across the junction, fluctuations in the field also modify the junction response. In traditional superconductor-insulator-superconductor (SIS) junctions, this can be described via the fluctuating phase difference $\phi(t)$ inserted in the dc Josephson relation $I_J=I_C \sin(\phi(t))$ within the resistively and capacitively shunted junction (RCSJ) model \cite{Tinkham}. As a result of these fluctuations, the junctions may switch to the dissipative state at bias currents lower than the critical current $I_C$, and therefore the measured critical current is often lower than its theoretical value that does not include the fluctuations. The difference between the two is proportional to the ratio of the temperature $T_{\rm env}$ describing the fluctuations and the Josephson energy $E_J=\hbar I_C/(2e)$ of the junction.

In contrast to such a simplified picture, fluctuations can nevertheless have a significant effect on the critical current even if the condition $E_J\gg{}k_BT$ is satisfied. Such an effect arises in superconductor--normal metal--superconductor (SNS) junctions, where the insulator is replaced by a normal metal (N) layer. These junctions also support finite supercurrents, but the effect of the electromagnetic field on this system is more complicated. In SNS junctions the supercurrent depends not only on the phase across the junction and its fluctuations, but also on the state of the electron system inside the normal metal \cite{heikkila02}. The latter is determined by a balance of energy currents between the electrons on the normal metal island and the other degrees of freedom in the system: phonons in the metal film, electrons inside the superconducting leads and --- via the fluctuations --- the electromagnetic environment of the junction (electron-photon coupling) \cite{ephotonrefs}. At temperatures low compared to the superconducting energy gap, Andreev reflection \cite{andreev} suppresses the energy transfer to the electrons in the leads. Therefore, the state of the electron system depends on the balance between electron-phonon and electron-photon coupling. 

In this Letter, we derive a linear-response collision integral describing electron-photon scattering in an SNS junction, and show that the effect of fluctuations on the SNS supercurrent is controlled by a parameter different from $k_BT_{\rm env}/E_J$, and that increasing the temperature of the environment can lead either to a decrease or an increase in the SNS supercurrent. The latter effect is directly related to the Eliashberg stimulated superconductivity \cite{eliashberg,Pauli} in SNS junctions. Besides changing the critical current, we show that this effect modifies the current-phase relation as the field absorption is greatly phase dependent.

\begin{figure}[tbhp]
\centering
\includegraphics[width=0.7\columnwidth]{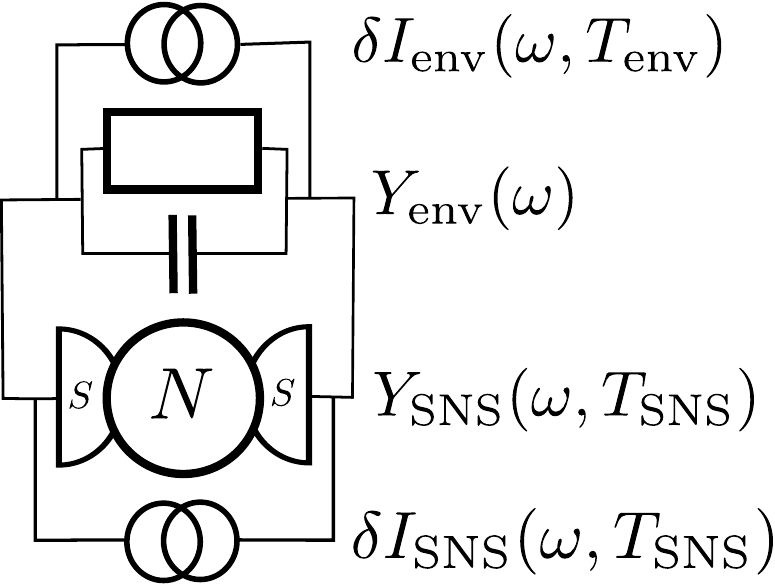}
\caption{Circuit model for an electrical environment with admittance $Y_{\rm env}$ coupling to a superconductor-normal metal-superconductor junction with admittance $Y_{\rm SNS}$.}
\label{fig:sketch}
\end{figure}

To specify our analysis, we consider the system depicted in Fig.~\ref{fig:sketch}. There, the SNS junction, described by its normal-state resistance $R_{\rm SNS}$, length $L$, and diffusion constant $D$, is coupled to an environment with admittance $Y_{\rm env}$. At first we consider this impedance as generic, but specify it more when discussing examples. Moreover, we assume that the SNS junction can be dc phase biased (e.g., in a SQUID setup) with phase $\phi$ and consider only the limit of a long junction, $L \gg \xi_0=\sqrt{\hbar D/(2 \Delta)}$, where $\Delta$ is the superconducting energy gap. The fluctuations related to the dissipative part of $Y_{\rm env}$ and those of the junction itself give rise to a fluctuating voltage $\Delta V(\omega)$ and a total fluctuating current $\Delta I(\omega)$ over the junction. The electron-photon coupling then results into a dissipated power $P_{e\gamma}=\langle \Delta V(\omega) \Delta I(\omega)\rangle$ into (or out of) the junction. The details of $\Delta V$ and $\Delta I$ are sensitive to the superconducting correlations in the SNS junction, which we take into account in the following.

First we note that the coupling of the electrons on the SNS junction to the electromagnetic environment can be envisaged as a photon exchange between two separate electron systems, described by energy distribution functions $f^{\rm SNS}_\epsilon$ and $f^{\rm env}_\epsilon$. Therefore, the collision integral for this process can be written in the form (below, $\hbar=k_B=1$)
\begin{align*}
I_{e\gamma}(\epsilon)&=\int d\omega d\epsilon' K(\omega,\epsilon,\epsilon')[f^{\rm SNS}_{\epsilon+\omega}f^{\rm env}_{\epsilon'-\omega}(1-f^{\rm SNS}_\epsilon)(1-f^{\rm env}_{\epsilon'})\\&-f^{\rm SNS}_{\epsilon}f^{\rm env}_{\epsilon'}(1-f^{\rm SNS}_{\epsilon+\omega})(1-f^{\rm env}_{\epsilon'-\omega})].
\end{align*}
Here the kernel $K(\omega,\epsilon,\epsilon')$ describes the coupling strength, and includes the effects of the superconducting correlations. We consider a macroscopic linear normal-metal noise source, for which the radiation absorption is energy independent, and which is in internal equilibrium, described by temperature $T_{\rm env}$. In that case the kernel does not depend on $\epsilon'$ and we can carry out the integral over $\epsilon'$ to get
\begin{equation}
\begin{split}
I_{e\gamma}(\epsilon) &= \int d\omega \omega K(\omega,\epsilon) [f^{\rm SNS}_{\epsilon+\omega}(1-f^{\rm SNS}_\epsilon)(n^{\rm env}_\omega+1)\\&-f^{\rm SNS}_\epsilon(1-f^{\rm SNS}_{\epsilon+\omega})n^{\rm env}_\omega],
\end{split}
\label{eq:electronphotoncollision}
\end{equation}
which describes electron-boson (photon) coupling. Here $n^{\rm env}_\omega$ is the Bose distribution function of the photons at temperature $T_{\rm env}$ and the two parts of the collision integral describe photon emission and absorption, respectively. 

We consider the effect of electron-photon interaction on the supercurrent flowing through the SNS junction at a certain phase difference $\phi$ across it,
\begin{equation}
 I_S(\phi)=\frac{1}{eR_{\rm SNS}}\int_0^\infty d\epsilon\; j_S(\epsilon,\phi)(1-2f(\epsilon)),
\end{equation}
where $j_S(\epsilon,\phi)={\rm Im}[j_E(\epsilon,\phi)]$ is the spectral supercurrent \cite{heikkila02} and $f(\epsilon)$ is the electron distribution function. In what follows, we consider linear response changes $\delta f$ of the distribution function due to the electron-photon coupling, and solve the kinetic equation
\begin{equation}
I_{e\gamma}(\epsilon)=I_{\rm eph}(\epsilon)=-\Gamma_{\rm eph} \nu(\epsilon) \delta f(\epsilon).
\label{eq:kineticequation}
\end{equation}
Here the collision integral $I_{\rm eph}(\epsilon)$ describing electron-phonon scattering is assumed to be the dominant source of energy relaxation. The latter form is valid in the linear response regime; $\nu(\epsilon)$ is the spatially averaged density of states inside the normal metal normalized to the normal state density of states at the Fermi level and $\Gamma_{\rm eph}$ is the electron-phonon scattering rate. Energy diffusion into the superconductors can be disregarded due to Andreev reflection \cite{andreev} when we consider energies much below the superconducting energy gap $\Delta$. Equation \eqref{eq:kineticequation} is therefore a valid approximation for long junctions $L \gg \xi_0$, where the relevant physics takes place around the Thouless energy $E_T=\hbar D/L^2$. 

In the linear response regime, the form of the kernel $K(\omega,\epsilon)$ in Eq.~\eqref{eq:electronphotoncollision} can be argued by considering the ac response of the junction \cite{Pauli2,Supp} to a fluctuating potential in the environment. We get $K(\omega,\epsilon)=K_{qp}(\omega,\epsilon)+K_{sc}(\omega,\epsilon)+K_{dy}(\omega,\epsilon)$, containing three parts due to quasiparticle, supercurrent, and dynamic responses on the ac potential. This yields
\begin{align}
&K(\omega,\epsilon)=\frac{4}{R_K \tau_D \omega^2} \frac{{\rm Re}(Y_{\rm env})}{|Y_{\rm env}+Y_{\rm SNS}|^2} \left\{\frac{1}{2}\langle[1+g(\epsilon)g(\epsilon+\omega)^*\right.\nonumber\\&+\frac{1}{2} F(\epsilon)F(\epsilon+\omega)^*+\frac{1}{2} \tilde F(\epsilon)\tilde F(\epsilon+\omega)^*]^{-1}\rangle^{-1}\nonumber\\
&   -\frac{1}{2}\partial_\phi {\rm Re}[j_E(\epsilon)+j_E(\epsilon+\omega)]\label{eq:kernel}\\&\left.-{\rm Im}\left[\frac{E_T}{2(\omega-2i\Gamma)}\frac{[j(\epsilon)-j(\epsilon+\omega)^*]^2}{\langle g(\epsilon)+g(\epsilon+\omega)^*\rangle}\right]\right\}\nonumber\\
&\equiv \frac{4}{R_K \tau_D \omega^2} \frac{{\rm Re}(Y_{\rm env})}{|Y_{\rm env}+Y_{\rm SNS}|^2} k(\epsilon,\omega,\phi),\nonumber
\end{align}
where $\langle \cdot \rangle$ denotes a spatial average over the normal metal island, $R_K=h/e^2$ is the resistance quantum, $\tau_D=L^2/D$ is the diffusion time, $Y_{\rm SNS}(\omega)$ is the admittance of the SNS junction \cite{Pauli2}, $g(\epsilon)$ and $F(\epsilon)$ are the normal and anomalous Green's functions inside the normal-metal island, $j_E(\epsilon)$ is the spectral supercurrent \cite{heikkila02}, and $E_T=D/L^2$ is the Thouless energy of the junction. These quantities can be calculated from the equilibrium Usadel \cite{usadel70,numericsnote} equation. 
Note that this approach disregards the equilibrium effect of phase fluctuations on the supercurrent \cite{Tinkham}. It is typically relevant when $T_{\rm env}$ is of the order $E_J$, or when $R_{||} \equiv {\rm Re}[Y_{\rm env}+Y_{\rm SNS}]^{-1}$ is of the order of $R_K$. In what follows, we thus assume $R_{||} \ll R_K$ and a large enough critical current to satisfy $T_{\rm env} \ll E_J$.

Below, we describe the external noise source by assuming it to consist of a resistance $R_{\rm env}$ in parallel with a capacitance $C$ (as in Fig.~\ref{fig:sketch}). The change in the supercurrent due to electron-photon coupling at linear response is then given by
\begin{align}
& \delta I_S(\phi)=\frac{1}{eR_{\rm SNS}}\frac{R_{\rm env}}{R_K \Gamma_{\rm e-ph} \tau_D}\int_{-\infty}^\infty d\epsilon\int_{-\infty}^\infty \frac{d\omega}{\omega}\frac{j_S(\epsilon,\phi)}{\nu(\epsilon,\phi)}\times\nonumber\\& \frac{k(\epsilon,\omega,\phi)}{\varphi(\tilde\omega,r,\omega_C)}{\rm sech}\left(\frac{\epsilon}{2T_{\rm SNS}}\right){\rm sech}\left(\frac{\epsilon+\omega}{2T_{\rm SNS}}\right)\sinh\left(\frac{\omega}{2T_{\rm SNS}}\right)\nonumber\\&\times \left[\coth\left(\frac{\omega}{2T_{\rm env}}\right)-\coth\left(\frac{\omega}{2T_{\rm SNS}}\right)\right],
\label{eq:dIS}
\end{align}
where the circuit parameters constitute a frequency-dependent term $\varphi=\left|1-i\omega/\omega_C+r\tilde y_{\rm SNS}(\omega)\right|^2$ describing the matching between the SNS junction and the environment, containing the parameters $r=R_{\rm env}/R_{\rm SNS}$, $\omega_C=1/(R_{\rm env} C)$ and $\tilde y_{\rm SNS} \equiv R_{\rm SNS} Y_{\rm SNS}$. The presence of a finite $\omega_C$ cuts the contribution from high frequencies --- if $\omega_C > \max\{T_{\rm SNS},T_{\rm env}\}$, the cutoff is provided by the temperature. In the opposite limit, we can expand the $\coth(\cdot)$ functions at low frequencies, and find that the effect is proportional simply to $T_{\rm env}-T_{\rm SNS}$.

From Eq.~\eqref{eq:dIS} we find that the overall magnitude of the change induced in the supercurrent by electron-photon coupling  is described by the parameter $\alpha \equiv R_{\rm env}/(R_K \Gamma_{\rm e-ph} \tau_D) (T_{\rm env}-T_{\rm SNS})/E_T=(R_{\rm env}/R_K) (T_{\rm env}-T_{\rm SNS})/(\Gamma_{\rm e-ph})$. The characteristics of the effect depend mostly on the following four parameters: temperature $T_{\rm SNS}$ of the SNS junction, phase $\phi$ across the junction, the charge relaxation rate $\omega_C\equiv (R_{\rm env}C)^{-1}$ and the matching factor $r\equiv R_{\rm env}/R_{\rm SNS}$. In the following, we analyze their effect in more detail.

\begin{figure}[h]
\centering
\includegraphics[width=\columnwidth]{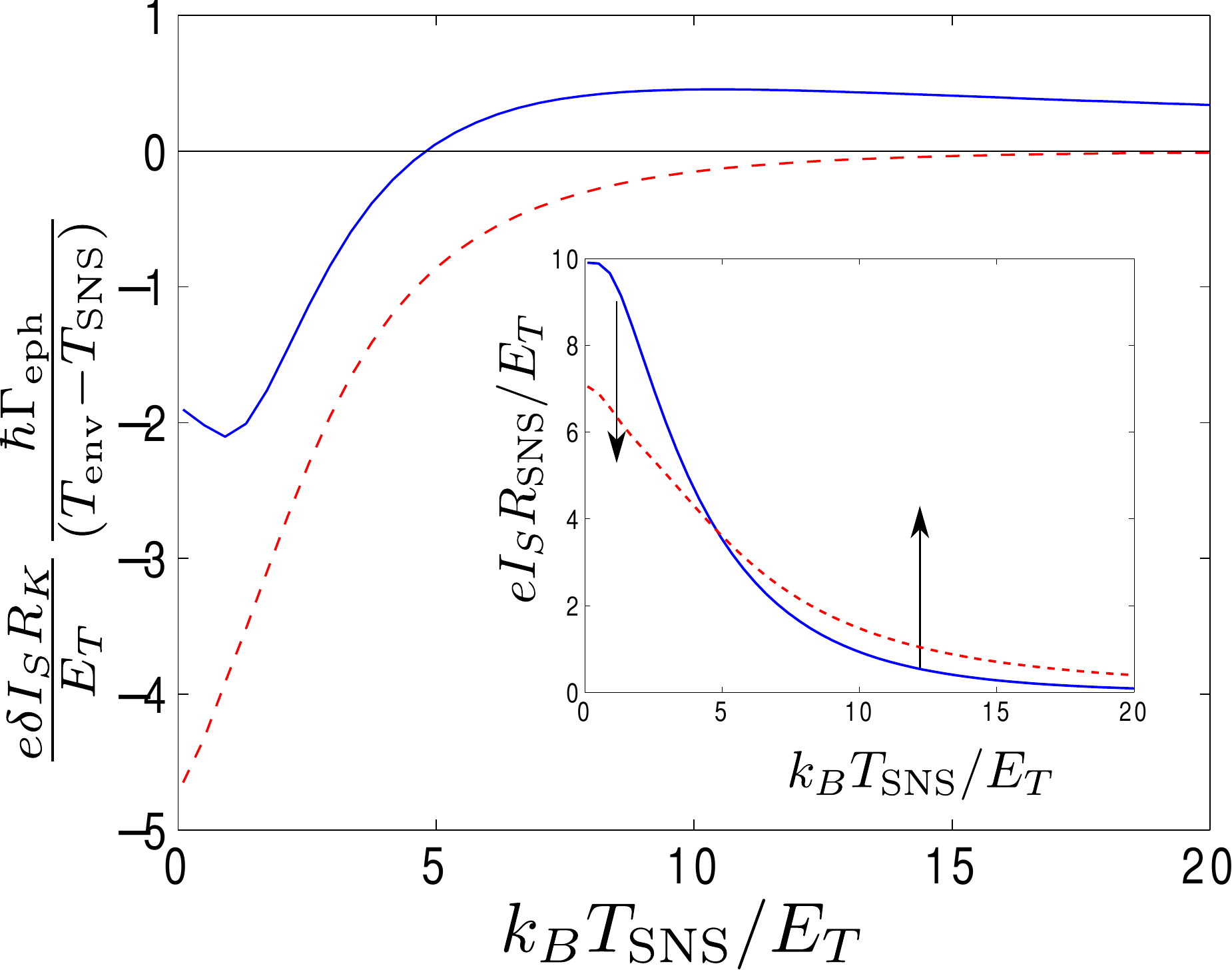}
\caption{(Color online): Electron-photon coupling induced change in the supercurrent vs. temperature of the SNS junction for $\phi=\pi/2$, $r=1$ and $\hbar \omega_C=5 E_T$. The blue solid line shows the result calculated with the coherent kernel $k(E,\phi)$ from Eq.~\eqref{eq:kernel} and the red dashed line the result that would be obtained in the incoherent limit where $k(E,\phi)=1$. Inset shows the total supercurrent in the absence of electron-photon scattering (blue solid line) and a sketch of the effect of electron-photon scattering with $T_{\rm env} > T_{\rm SNS}$ (red dashed line). The arrows point the direction of the change in the supercurrent as the noise temperature of the environment is increased. Strictly speaking, the dashed line is for $k_B T \lesssim 5E_T$ outside of the linear response regime, but it captures the qualitative effect correctly. Note that in practice when considering $\delta I_S(T)$, one should take into account the temperature dependence of the electron-phonon scattering $\Gamma_{\rm eph}\propto T_{\rm SNS}^3$.}
\label{fig:Tdep}
\end{figure}

The effect of electron-photon coupling on the supercurrent at phase $\phi=\pi/2$ (close to the phase giving the maximum supercurrent) as a function of the temperature $T_{\rm SNS}$ of the phonons in the SNS junction is depicted in Fig.~\ref{fig:Tdep}. The inset shows the overall supercurrent as a function of temperature in the presence and absence \cite{heikkila02,dubos01} of the electron-photon coupling (corresponding, hence, to the cases $T_{\rm env} > T_{\rm SNS}$ and $T_{\rm env}=T_{\rm SNS}$, respectively). We find out that at low $T_{\rm SNS}$, the supercurrent decreases as the SNS junction heats up due to the absorption of power from the electromagnetic environment. However, at higher temperatures, $k_B T_{\rm SNS} \gtrsim 5 E_T$, the electron-photon coupling to a high-temperature noise source leads to an {\it increase} in the supercurrent. This is a true nonequilibrium effect and resembles the stimulation of superconductivity encountered also in the presence of monochromatic driving of the junction \cite{Pauli,warlaumont79}. Note that this happens at the linear response of the junction to the electron-photon coupling: increasing $T_{\rm env}$ further eventually leads to a decrease of the overall supercurrent. 

\begin{figure}[h]
\centering
\includegraphics[width=\columnwidth]{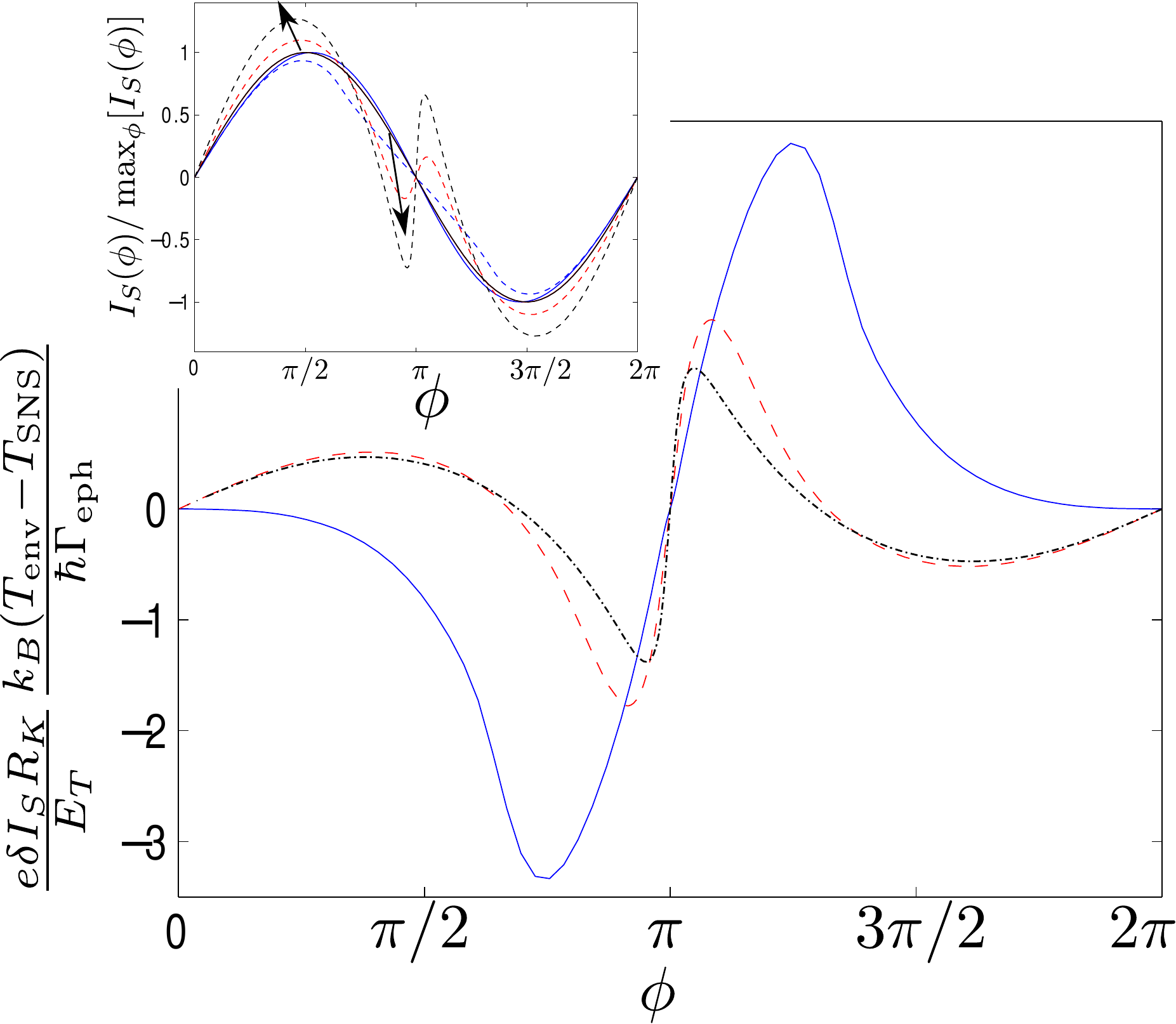}
\caption{(Color online): Electron-photon coupling induced change in the supercurrent vs. phase $\phi$ with $r=1$ and $\hbar \omega_C=5 E_T$ and three temperatures $k_B T_{\rm SNS}$: $3 E_T$ (blue solid line), $10 E_T$ (red dashed line) and $15 E_T$ (black dash-dotted line). Inset shows the normalized current-phase relation $I_S(\phi)/\max_\phi I_S(\phi)$ in the absence of electron-photon scattering (solid lines) and a sketch of the effect of electron-photon scattering with $T_{\rm env} > T_{\rm SNS}$ (dashed lines) at the same three temperatures. There, the arrows point the direction of the supercurrent change as $T_{\rm env}$ is increased.}
\label{fig:phidep}
\end{figure}

The strongest enhancement of the supercurrent can be found for phases around $\phi \approx \pi/2$. This effect can be traced to the existence of a minigap of size $\sim E_T$ in the excitation spectrum (and the kernel $k(E,\phi)$). On the other hand, for phases $\phi \approx \pi$, the minigap closes and the electron-photon coupling only suppresses the supercurrent. This characteristics is shown in Fig.~\ref{fig:phidep}, which shows the supercurrent change $\delta I_S$ as a function of the phase. A similar shape of the current-phase relation has been found for monochromatic driving, both theoretically \cite{Pauli} and experimentally \cite{fuechsle09}. 

\begin{figure}[h]
\centering
\includegraphics[width=\columnwidth]{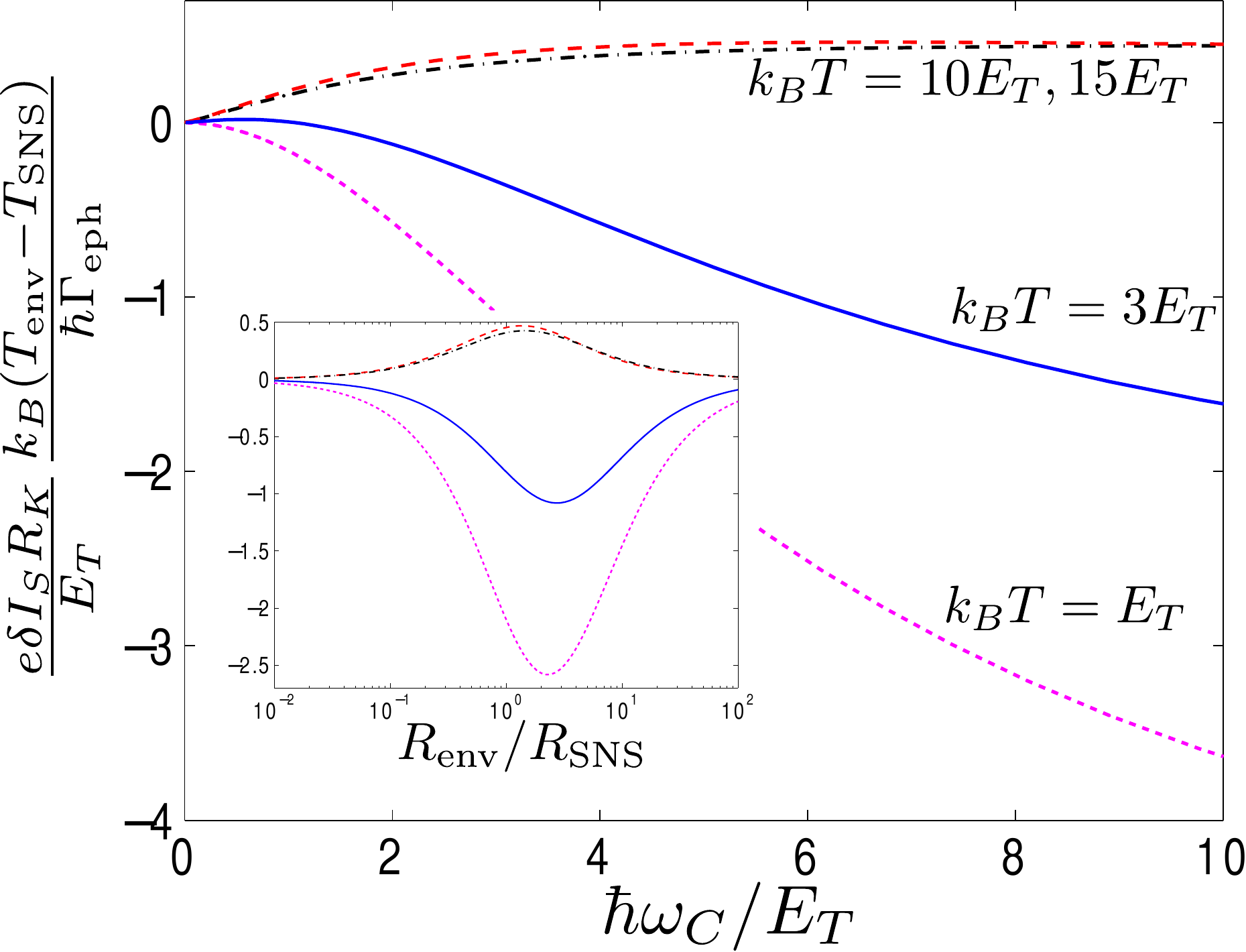}
\caption{(Color online): Electron-photon coupling induced change in the supercurrent at $\phi=\pi/2$ vs. the parameters of the circuit at the temperatures indicated in the figure. Main figure: effect of the changing charge relaxation rate $\omega_C=1/(R_{\rm env} C)$ acting as an effective high-frequency cutoff on the electron-photon coupling. The curves have been calculated with $R_{\rm env}=R_{\rm SNS}$. Inset shows the effect of changing the ratio $R_{\rm env}/R_{\rm SNS}$ while keeping $\hbar\omega_C=5E_T$ (note the logarithmic scale on the horizontal axis). The two figures show the same quantity (with the same scaling).}
\label{fig:matching}
\end{figure}

The effect of electron-photon coupling is naturally strongest when the resistance describing the electromagnetic environment equals the SNS normal-state resistance, i.e., $r=R_{\rm env}/R_{\rm SNS} \approx 1$, and as much noise as possible is coupled to the junction, and therefore $\omega_C$ is as large as possible \cite{frequencynote}. For completeness, we show the effect of varying these parameters in Fig.~\ref{fig:matching}. We find out that the major effect on the current increase comes from frequencies $\omega \approx E_T$, so that a further increase of the cutoff frequency beyond a few $E_T$ does not affect the increase much. On the other hand, the incoherent reduction of the supercurrent (at low temperatures, for example) increases in strength as the noise bandwidth is increased. We also point out that the ``optimal'' matching of noise takes place at $R_{\rm env}$ somewhat larger than the normal-state resistance $R_{\rm SNS}$ of the SNS junction, but the order of magnitude of the effect depends quite weakly on their ratio.

Let us estimate the typical parameters for the electron-photon coupling in SNS junctions \cite{dubos01}, where the authors report at low temperatures a 7 \% difference between their experimental results and the theory that does not take into account phase fluctuations. A Cu wire of length 1 $\mu$m, diffusion constant $D=0.02$ m$^2$/s and normal-state resistance $R_{\rm SNS} =0.2$ $\Omega$ has a Thouless energy $E_T \approx 13$ $\mu$eV and a zero-temperature critical current of 650 $\mu$A. This corresponds to the Josephson energy $E_J=\hbar I_C/(2e) =1.3$ eV, allowing to increase the (noise) temperature of the electromagnetic environment to very large values before any phase diffusion could be observed. Increasing $T_{\rm SNS}$ decreases the critical current and thereby $E_J$, but the observation of phase diffusion would require quite high $T_{\rm SNS}$. On the other hand, the change in the supercurrent due to electron-photon coupling is $\delta I_S/I_S\sim aR_{\rm SNS}^2k_B (T_{\rm env}-T_{\rm SNS})/(cR_K R_{\rm env}\hbar \Gamma_{\rm eph})$, where we assume $R_{\rm SNS}<R_{\rm env}$, $a$ is the dimensionless number plotted in Figs.~\ref{fig:Tdep}-\ref{fig:matching} at perfect matching, and $c=eI_SR_{\rm SNS}/E_T\approx 10$ at $T_{\rm SNS}\lesssim E_T/k_B$. For Cu, a typical electron-phonon scattering rate at $T=100$ mK is 20 kHz \cite{giazotto06}, corresponding to the temperature scale of $\hbar \Gamma_{\rm eph}/k_B \approx 0.15$ $\mu$K. Therefore, for a typical $R_{\rm env} = 50$~$\Omega$, we get $\delta I_S/I_S=a/c\sim 0.07$ for $T_{\rm env}-T_{\rm SNS}=8$ K. As the SNS junctions are typically connected to a measurement equipment residing at higher temperatures, the noise coupling from such equipment may well result in noise temperatures of this order of magnitude. Moreover, many experiments are conducted on higher-resistance samples than those considered above, in which case the required temperature difference decreases. Therefore, our results may explain the typically encountered difference between the experimental results and the standard theoretical predictions \cite{otherexp} as being caused by electron-photon coupling. However, to really probe the effect we are predicting, the environmental noise should be systematically varied while measuring the supercurrent. The previous can be done for example by passing a large heating current through a macroscopic shunt resistor of the SNS junction.

{\em Conclusions.} We have shown that whereas the typical and well-known mechanism of the effect of phase fluctuations on the supercurrent through superconductor-normal-metal-superconductor junctions, dependent on the parameter $k_B T/E_J$, can often be disregarded, the heat current due to the temperature difference between the electromagnetic environment and the SNS junction leads to much more pronounced effects. At low temperatures $k_B T_{\rm SNS} \lesssim 5E_T$, this results into a suppression of the observed supercurrent, but what is more remarkable, for $k_B T_{\rm SNS} \gtrsim 5E_T$, we predict an {\it increased} supercurrent, competing with the exponentially suppressed bare supercurrent. Our predictions should be tested by simply varying the temperature of the electromagnetic environment while keeping that of the SNS junction constant. Besides weak links fabricated of normal metals, similar effects can be expected for other types of conducting weak links, such as those made of graphene, carbon nanotubes or semiconductor nanowires.

\begin{acknowledgments}
We thank M.A. Laakso, J.C. Cuevas and F.S. Bergeret for discussions. This work was supported by the Finnish Foundation for Technology Promotion, the Academy of Finland and the European Research Council (Grant No. 240362), and the Emmy-Noether program of the Deutsche Forschungsgemeinschaft.
\end{acknowledgments}

\begin{appendix}

\begin{widetext}
\section{Appendix: Environment-controlled change in the current}

In this supplementary material, we give details on how the effect of
fluctuations on a superconductor--normal metal--superconductor
junction can be derived from microscopic theory.

We describe the effect of fluctuations by considering the Keldysh path
integral action \cite{kindermann2003-dvf} of the circuit of
Fig.~\ref{fig:environment}:
\begin{align}
  S[\Phi,\chi] = S_{\rm SNS}[\Phi+\chi] + S_{\rm env}[\Phi]
  \,,
\end{align}
where $\Phi(t)=\frac{e}{\hbar}\int^t \dd{t}V(t)$ is the
electromagnetic phase drop across the SNS junction, with the quantum
and classical components $\Phi^{cl/q}(t)=(\Phi_+(t)\pm\Phi_-(t))/2$
related to its values $\Phi_\pm$ on the two Keldysh branches. We also
add a generating field $\chi$, so that the current in the SNS can be
written as
\begin{align}
  \label{eq:current}
  I(t)
  =
  \int\DD{\Phi}e^{iS_{\rm env}[\Phi] + iS_{\rm SNS}[\Phi]}
  \frac{\delta S_{\rm SNS}[\Phi+\chi]}{\delta \chi^q(t)}\rvert_{\chi=0}
  \,.
\end{align}
We assume the environment is characterized by an admittance $Y_{\rm
  env}$ describing a circuit element at equilibrium.  We also assume
that the saddle point of the action corresponds to a constant
superconducting phase difference $\varphi_0$ over the junctions,
corresponding to a dc supercurrent $I_0$ through the SNS. In terms of
fluctuations $\phi=\Phi-\Phi_0$ around the saddle point $\Phi_0$, the
environment action can be written as:

\begin{align}
  \label{eq:environment-action}
  S_{\rm env}[\phi]
  &=
  -
  I_0
  \int_{-\infty}^\infty\dd{t}
  \phi^q(t)
  +
  \int_{-\infty}^\infty
  \frac{\dd{\omega}}{2\pi}
  \begin{pmatrix}
    \phi^{cl}(\omega) \\ \phi^q(\omega)
  \end{pmatrix}^\dagger
  \begin{pmatrix}
    0 & [i\omega Y_{\rm env}(\omega)]^* \\
    i\omega Y_{\rm env}(\omega) & 
    2i\omega\coth(\frac{\omega}{2T_{\rm env}})\Re Y_{\rm env}(\omega)
  \end{pmatrix}
  \begin{pmatrix} 
    \phi^{cl}(\omega) \\ \phi^q(\omega)
  \end{pmatrix}
  \,,
\end{align}
which produces the correlators $\avg{\phi\phi}$ expected of a
classical circuit element. Here and below, we use natural units in
which $e=\hbar=k_B=1$.

\begin{figure}
  \includegraphics{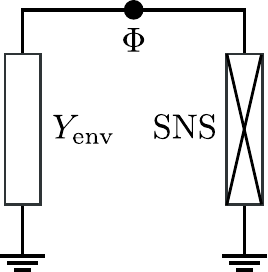}
  \caption{\label{fig:environment}
    SNS junction and its electromagnetic environment.
    $\Phi$ is the electromagnetic phase across both elements.
  }
\end{figure}

Consider now the action of the SNS junction similarly expanded in
fluctuations:
\begin{align}
  \label{eq:sns-full-action}
  S_{\rm SNS}[\phi,0] 
  &=
  I_{0}
  \int_{-\infty}^\infty\dd{t}
  \phi^q(t)
  +
  \int_{-\infty}^\infty
  \frac{\dd{\omega}}{2\pi}
  \begin{pmatrix}
    \phi^{cl}(\omega) \\ \phi^q(\omega)
  \end{pmatrix}^\dagger
%  \begin{pmatrix}
%    0 & {\cal V}^A(t-t') \\
%    {\cal V}^R(t-t') & {\cal V}^K(t-t')
%  \end{pmatrix}
  \begin{pmatrix}
    0 & [i\omega Y_{\rm SNS}(\omega)]^* \\
    i\omega Y_{\rm SNS}(\omega) & 
    2i\omega\coth(\frac{\omega}{2T_{SNS}})\Re Y_{\rm SNS}(\omega)
  \end{pmatrix}
  \begin{pmatrix} 
    \phi^{cl}(\omega) \\ \phi^q(\omega)
  \end{pmatrix}
  +
  A[\phi]
  \\
  &=
  S_{SNS,0}[\phi] + A[\phi]
  \,,
\end{align}
where $I_{0}=I_{\rm eq}(\varphi_0)$ is the equilibrium supercurrent.
The form of the second-order term is fixed by the fact that it
describes the linear response of the SNS junction around
equilibrium. It is similar to Eq.~\eqref{eq:environment-action}, but
with the admittance (for which approximations are
known\cite{virtanen2011-lar}) and temperature replaced by those of the
SNS junction.  The term $A$ describes higher-order corrections to the
behavior of the SNS due to the fluctuations. When $T_{\rm
  env}\ne{}T_{SNS}$, part of these corrections comes from
nonequilibrium associated with the energy transfer from one subsystem
to the other by phase fluctuations.

The next step would be to compute $S_{\rm SNS}[\phi]$ based on a
microscopic model. This problem is however equivalent to finding the
full counting statistics \cite{Note1} of the SNS junction under a
general time-dependent drive, which for long junctions is a difficult
problem.  Below, we argue that nevertheless, in the limit of small
phase fluctuations, the physics we are interested in here is described
by the response of the junction to \emph{classical} fluctuations.

We first expand the higher-order SNS part $A$ in
Eq.~\eqref{eq:current} in $\phi$, and obtain:
\begin{align}
  I(t)
  &=
  \int\DD{\phi}
  e^{i S_{\rm env}[\phi] + iS_{SNS,0}[\phi]}
  \left\{
  I_{0}
  +
  2
  \int_{-\infty}^\infty\dd{t'}
  {\cal V}^R(t-t')\phi^{cl}(t')
  +
  \frac{\delta \tilde{A}[\phi+\chi]}{\delta\chi^q(t)}\rvert_{\chi=0}
  +
  \ldots
  \right\}
  \\
  \label{eq:current-expansion}
  &=
  \avg{
    \frac{\delta S_{\rm SNS}[\phi+\chi]}{\delta\chi^q(t)}\rvert_{\chi=0}
  }_\phi
  +
  {\cal O}(\phi^3)
  =
  \avg{
    I_{SNS}[\phi]
  }_\phi
  +
  {\cal O}(\phi^3)
  \,,
\end{align}
where $\tilde{A}$ contains only the third-order terms, ${\cal V}^R$ is
the Fourier transform of $i\omega Y_{\rm SNS}(\omega)$, and the
averages are computed with the quadratic part of the action.  The
second term on the first line vanishes, $\avg{\phi}_\phi=0$, but the
third is finite. Note the structure of this approach: one first
computes the current $I_{SNS}[\phi]$ through the junction using a
fixed time dependence of the phase fluctuation $\phi$, and finally
averages the result over Gaussian fluctuations as determined by the
admittances.

We now observe the following: Eq.~\eqref{eq:environment-action}
implies that the temperature of the environment $T_{\rm env}$ appears
in Eq.~\eqref{eq:current-expansion} only in correlation functions
$\avg{\phi^{cl}\phi^{cl}}_\phi$.  Therefore, if we consider only the
effect of $T_{\rm env}$ on the current, we find that in the leading
order in the phase fluctuations, the change in the current due to
$T_{\rm env}\ne{}T_{SNS}$ is
\begin{align}
  \label{eq:current-T-dependence}
  \delta I(t)
  \equiv
  I(t) - I(t)\rvert_{T_{\rm env}=T_{SNS}}
  =
  \avg{
    I_{SNS}[\phi]
  }_{\phi^{cl}}
  -  
  \avg{
    I_{SNS}[\phi]
  }_{\phi^{cl}}
  \rvert_{T_{\rm env}=T_{SNS}}
  \,,
\end{align}
where the field averages are taken considering $\phi$ as a classical
field, $\phi^q=0$. This observation considerably simplifies the
approach: we can first compute the current for a given time dependence
of a \emph{classical} phase difference over the junction, and then
average the result over Gaussian fluctuations. The effect of such
classical fluctuations on the supercurrent can be obtained as an
extension of our earlier results
\cite{virtanen2010-tom,virtanen2011-lar} for the effect of a
monochromatic classical drive. This is outlined in the next section.

We now comment on how small the phase fluctuations must be for the validity of our model. The
criterion is that truncating the expansion
Eq.~\eqref{eq:current-expansion} must remain accurate.  The first
requirement is that the average phase fluctuations should be small,
$\avg{\phi(t)\phi(0)}\ll{}1$. Assuming total parallel admittance
$Y=R^{-1} + 1/(i\omega L)$ of the SNS junction and the environment,
this is equivalent to the restrictions $R\ll{}R_K$ and $L k_B T/(\hbar
R_K)\ll1$. The former is satisfied for typical SNS junctions. If the
inductance comes from the Josephson inductance of the SNS junction,
the latter is equivalent to $E_J\gg{}k_BT$, which is the typical
condition for fluctuations to have a small effect. There is also a
requirement that the nonequilibrium corrections to the SNS current are
small enough to remain in the linear regime. As noted in the main
text, this condition can be written as $R/R_K\times{} k_B (T_{\rm
  env}-T_{SNS})\ll{}\Gamma_{\rm e-ph}$, where $\Gamma_{\rm e-ph}$ is
the electron-phonon relaxation rate, which should dominate energy
relaxation inside the SNS junction.

\section{Effect of classical phase fluctuations}

The effect of small classical phase fluctuations on the dc current in
a SNS junction can be studied by expanding the time-dependent Usadel
equation \cite{usadel1970-gde,larkin1986-ns} in the fluctuating
electric field associated with the time-dependent phase difference
$\phi(t)$. Such a calculation was done in
Ref.~\onlinecite{virtanen2010-tom} for a monochromatic excitation
$\phi(t)=\phi_0\cos(\omega_0 t)$.  A kinetic equation for an arbitrary
small perturbation $\phi(t)$ can however also be derived following the
same steps. Our starting point here is the kinetic equation for the dc
component of the electron distribution obtained in
Ref.~\onlinecite{virtanen2010-tom}, which does not make assumptions
about the time-dependence of the small perturbation:
\begin{align}
  \label{eq:kinetic-equation}
  8 \Gamma_{\rm e-ph}\overline{\nu(\epsilon)}
  [h(\epsilon,\epsilon+\omega) - h_0(\epsilon,\epsilon+\omega)]
  =
  \tr
  [-iA\tau_3, \hat{j}^K]_\circ(\epsilon,\epsilon+\omega)
  +
  {\cal O}(A^3)
  \,,
\end{align}
where $\omega\to0$, and the commutator involves a convolution over
energy arguments. Here, $h$ is the energy mode (longitudinal) electron distribution function, $\hat{j}^K \equiv (\check g \circ \check \nabla \check g)^K$ is the current related to the Keldysh Green's function $\check g$, and $\check \nabla$ is the gauge-invariant gradient. In particular, the charge current is proportional to $\tr \hat{\tau}_3 \hat{j}^K$. Moreover,
$A(\omega,\omega')=\phi(\omega-\omega')/L$ is the Fourier-transformed
vector potential corresponding to a constant electric field associated
with the fluctuation $\phi$ in a junction of length $L$,
$\overline{\nu(\epsilon)}$ the position-averaged density of states in
the absence of fluctuations, and
$h_0(\epsilon,\epsilon')=2\pi\delta(\epsilon-\epsilon')h_0(\epsilon)$,
$h_0(\epsilon)=\tanh\bigl(\frac{\epsilon}{2T_{SNS}}\bigr)$. 

Averaging Eq.~\eqref{eq:kinetic-equation} over the fluctuating fields,
we find:
\begin{align}
  8 \Gamma_{\rm e-ph}\overline{\nu(\epsilon)}
  [\avg{h(\epsilon,\epsilon')}_\phi - h_0(\epsilon,\epsilon')]
  =
  -i\int_{-\infty}^\infty\frac{\dd{\epsilon_1}}{2\pi}
  \avg{
  A(\epsilon-\epsilon_1)\tr\hat{\tau}_3\hat{j}^K(\epsilon_1,\epsilon')
  -
  A(\epsilon_1-\epsilon')\tr\hat{\tau}_3\hat{j}^K(\epsilon,\epsilon_1)
  }_\phi
  \,.
\end{align}
In Ref.~\onlinecite{virtanen2011-lar} we showed that in linear order
in the field, the quantity $\tr\hat{\tau}_3\hat{j}^K$ can be
approximated by
\begin{align}
  \tr\hat{\tau}_3\hat{j}^K(\epsilon,\epsilon')
  \simeq
  \tr\hat{\tau}_3\hat{j}^K_{\rm eq}(\epsilon,\epsilon')
  +
  A(\epsilon-\epsilon')
  M(\epsilon,\epsilon')
  \,,
\end{align}
with a known linear response coefficient
$M(\epsilon,\epsilon')=M(\epsilon',\epsilon)^*$.  Combining this
result with the kinetic equation, we find
\begin{align}
  8 \Gamma_{\rm e-ph}\overline{\nu(\epsilon)}
  [
  h(\epsilon)
  -
  h_0(\epsilon)
  ]
  &=
  2
  \int_{-\infty}^\infty\frac{\dd{\omega}}{2\pi}
  S_\phi(\omega, T_{\rm env}, T_{SNS})
  L^{-2}
  \Im
  M(\epsilon,\epsilon+\omega) 
  \\
  &=
  \pi^{-1}
  \int_{-\infty}^\infty\dd{\omega}
  S_\phi(\omega, T_{\rm env}, T_{SNS})
  \tau_D^{-1}
  2
  k(\omega,\epsilon)[h_0(\epsilon)-h_0(\epsilon+\omega)]
  \,,
\end{align}
where
$\avg{h(\epsilon,\epsilon')}_\phi=2\pi\delta(\epsilon-\epsilon')h(\epsilon)$,
the factor $k(\omega,\epsilon)$ is defined in Eq.~(4) in the main
text, and
\begin{align}
  S_\phi(\omega, T_{\rm env}, T_{SNS})
  =
  4\pi
  \frac{
    \Re[Y_{\rm env}(\omega)]\coth\bigl(\frac{\omega}{2T_{\rm env}}\bigr)
    +
    \Re[Y_{\rm SNS}(\omega)]\coth\bigl(\frac{\omega}{2T_{SNS}}\bigr)
  }{\omega R_K |Y_{\rm env}(\omega) + Y_{\rm SNS}(\omega)|^2}
  \,,
\end{align}
is the symmetrized phase fluctuation spectrum from the
field correlators,
$\avg{A(\omega)A(\omega')}_\phi=L^{-2}\avg{\phi^{cl}(\omega)\phi^{cl}(\omega')}=L^{-2}2\pi\delta(\omega+\omega')S(\omega,T_{\rm
  env},T_{SNS})$. Here, $R_K=h/e^2=2\pi$ in natural units.

As we argued in Ref.~\onlinecite{virtanen2010-tom}, when the
electron-phonon relaxation is small compared to the inverse dwell time
in the junction, $\Gamma_{\rm e-ph}\ll{}E_T=\hbar D/L^2$, the change
in the supercurrent through the junction is mainly determined by the
change in the distribution function.  Applying now the result in
Eq.~\eqref{eq:current-T-dependence} gives
\begin{align}
  \delta I
  &=
  \frac{1}{eR_{SNS}}
  \int_{-\infty}^\infty \dd{\epsilon} j_S(\epsilon) \delta h(\epsilon)
  \,,
  \\
  \Gamma_{\rm e-ph}\overline{\nu(\epsilon)}
  \delta h(\epsilon)
  &=
  \Gamma_{\rm e-ph}\overline{\nu(\epsilon)}
  [h(\epsilon) - h(\epsilon)\rvert_{T_{\rm env}=T_{SNS}}]
  \\
  &=
  \frac{1}{4\pi}
  \int_{-\infty}^\infty\dd{\omega}
  [S_\phi(\omega, T_{\rm env},T_{SNS}) - S_\phi(\omega, T_{SNS},T_{SNS})]
  \tau_D^{-1}
  k(\epsilon,\omega)
  [h_0(\epsilon) - h_0(\epsilon+\omega)]
  \\
  &=
  \frac{1}{4}
  \int_{-\infty}^\infty\dd{\omega}
  \omega
  K(\omega,\epsilon)
  \Bigl[\coth\Bigl(\frac{\omega}{2T_{\rm env}}\Bigr) - 
  \coth\Bigl(\frac{\omega}{2T_{SNS}}\Bigr)\Bigr]
  [h_0(\epsilon) - h_0(\epsilon+\omega)]
  \\
  &=
  \int_{-\infty}^\infty\dd{\omega}
  \omega
  K(\omega,\epsilon)
  [
  f_{\epsilon+\omega}(1 - f_\epsilon)(n_\omega^{\rm env}+1) 
  -
  f_\epsilon(1 - f_{\epsilon+\omega})n_\omega^{\rm env}
  ]
  \,,
\end{align}
where $f_\epsilon=\frac{1-h_0(\epsilon)}{2}$ is the equilibrium Fermi
function, and $n_\omega^{\rm env}=[e^{\omega/T_{\rm env}}-1]^{-1}$ the
Bose function.  We therefore find that the change in the current is
determined by an electron-boson collision integral, and we obtain the
kernel $K(\omega,\epsilon)$ given in Eq.~(4) of the main text. The
approach we used to derive this result here, however, is restricted to
the leading order in the field amplitude and small nonequilibrium
effects.

\end{widetext}
\end{appendix}


\begin{thebibliography}{99}
\bibitem{frequencynote} The theory detailed in \cite{Pauli2} leading to Eq.~\eqref{eq:kernel} assumes a position independent vector potential inside the junction. This assumption breaks down at high frequencies, $\hbar\omega \gtrsim 10 E_T$, and therefore we limit ourselves to cutoff frequencies below this scale. We do not expect significant qualitative changes to the collision integral even beyond this point, but quantitative details are altered.
\end{thebibliography}
\end{document}